# Assessing the Lognormal Distribution Assumption For the Crude Odds Ratio: Implications For Point and Interval Estimation

David D. Newstein MS, DrPH, Hunter College, CUNY, Department of Mathematics and Statistics

## Summary

**Background:** The assumption that the sampling distribution of the crude odds ratio ($OR_{crude}$) is a lognormal distribution with parameters $\mu$ and $\sigma$ leads to the incorrect conclusion that the expectation of the log of $OR_{crude}$ is equal to the parameter $\mu$. In fact, $\exp(\mu)$ is the median of the lognormal distribution not the mean. If a different parameter is obtained as the expected value of this distribution, $\exp(\mu^*) = \exp(\mu + \sigma^2/2)$ then this quantity can be used to obtain a new estimate of $OR_{true}$. Thus, if we define $OR^*$ by $\ln(OR^*) = \ln(OR_{crude}) - \sigma^2/2$, then $OR^*$ may yield a less biased estimate of $OR_{true}$ than $OR_{crude}$.

**Methods**: The standard method of point and interval estimation for the odds ratio (I), was compared with a modified method utilizing $OR^*$ instead of $OR_{crude}$. Confidence intervals were obtained utilizing $\mu^* = \ln(OR^*)$ and $\sigma$ by both parametric bootstrap simulations with a percentile derived confidence interval (II), and a simple calculation done by replacing $\ln(OR_{crude})$ with $\ln(OR^*)$ in the standard formula (III). These two methods have been shown to be mathematically equivalent. Monte Carlo (MC) simulations were conducted to compare these three methods of point and interval estimation along with a method proposed by Barendregt (IV), who noted the bias present in the standard estimate of $\mu$ by $\ln(OR_{crude})$.

**Results:** In simulations, the estimation methods II and III yielded coverage probabilities closest to $1 - \alpha = .95$. Also, as demonstrated by the MC simulations, these two methods exhibited the least biased point estimates and the narrowest confidence intervals of the four estimation approaches.

**Conclusions:** Monte Carlo simulations prove useful in validating the inferential procedures used in data analysis. In the case of the odds ratio, the standard method of point and interval estimation is based on the assumption that the crude odds ratio has a sampling distribution that is lognormal. Utilizing this assumption, as well as the formula for the expectation of this distribution function, an alternative estimation method was obtained for $OR_{true}$ (but different from a method from the earlier report), that yielded point and interval estimates that MC simulations indicate are the most statistically valid.

**Background:**

Relative effect measures have been utilized to interpret and report the magnitude of benefits or risks to human health of diverse exposures since the 1950s. These measures are estimated from data from both observational and experimental studies. Additionally, confidence intervals are typically calculated utilizing these point estimates along with estimates of their standard errors. Hypothesis tests of the null hypothesis test that the true value of the effect is equal to one (ie. there is no effect of exposure on outcome). These tests are conducted utilizing the same point estimates along with the same estimates of their standard errors as are used to derive the confidence intervals. Mathematically, the typical assumption for deriving interval estimates for the true value of the effect, as well as hypothesis tests, is to assume that the sampling distributions of the estimates are lognormal probability distributions. This report will concentrate on the theory of the standard method of point and interval estimation for the odds ratio, although the theory may be applied to all commonly used relative effect measure estimates (ie. the incidence density ratio and the cumulative risk ratio). Monte Carlo simulations of a simple clinical experiment are used to assess statistical conclusion validity of the confidence intervals for the standard and three alternative methods and the associated tests of the null hypothesis that the true odds ratio is equal to one for each of the four methods. The statistical conclusion validity of the interval estimates are assessed by the simulated empirical coverage probability (for a given $OR_{true}$). The validity of the hypothesis tests are assessed by examining the simulated empirical power for rejecting the null hypothesis that the true odds ratio is equal to one when the alternative hypothesis is true (ie. the true odds ratio is **not** equal to one). Comparisons are made between the standard method of estimation and the three alternative estimation approaches which may yield less biased point estimates along with better interval estimates and hypothesis tests.

The odds ratio is a commonly employed relative effect measure in epidemiologic research. Its widespread use is mainly a result of its application to multiple logistic regression modelling as well as its use in the analysis of data from case control studies. The crude odds ratio ($OR_{crude}$ or unadjusted sample OR) may be calculated from the standard two by two table of disease versus exposure by the familiar "cross product" formula [1]. This crude measure often serves as an estimate for the unknown true odds ratio ($OR_{true}$).

Common practice in both hypothesis testing and interval estimation for the odds ratio is to log transform $\widehat{OR}_{crude}$ and then obtain an estimate for the standard deviation of $ln(\widehat{OR}_{crude})$ utilizing the assumption that $ln(\widehat{OR}_{crude})$ is normally distributed. This assumption is mathematically equivalent to assuming that $\widehat{OR}_{crude}$ is lognormally distributed. The assumption that $ln(\widehat{OR}_{crude})$ is normally distributed is usually justified by an asymptotic result known as the Delta Method [2]. This theorem asserts that if a real valued function of a random vector $x_n$, $say\ f(x_n)$, is asymptotically normally distributed with mean u and standard deviation $\sigma_n$ (where $\lim_{n\to\infty} \sigma_n = 0$), then for any real valued function g, which has a continuous first derivative at u: $\frac{g(f(x_n))-g(u)}{g'(u)\sigma_n} \to N(0,1)\ as\ n \to \infty$. For the crude odds ratio $x_n = (a_n, b_n, c_n, d_n)$ which are the cell frequencies of the 2x2 table for a sample of size n, $f(x_n) = \frac{a_n \cdot d_n}{b_n \cdot c_n}$, and g(y) = log(y).

For the odds ratio the Delta Method yields an estimate for $\sigma$ which is derived from a Taylor Series expansion of $g(f(x)) = ln(\widehat{OR}_{crude})$. In practical applications, only the constant and linear terms of the expansion are evaluated. This is how the formula for the common estimate for the standard deviation of $ln(\widehat{OR}_{crude})$ is derived.

The standard method of calculation of the confidence interval for the true odds ratio equates $ln(\widehat{OR}_{crude})$ with an estimate of the parameter $\mu$ from a normal distribution (the assumed sampling distribution of $ln(\widehat{OR}_{crude})$). A problem arises because the above assumptions imply that the expectation of $ln(\widehat{OR}_{crude})$ is **not** equal to the parameter $\mu$ (ie. $E(ln(\widehat{OR}_{crude})) \neq \mu$). This is a result of the relationship between the parameters when the parametrization for the lognormal distribution arises from an exponentiated normally distributed random variable with the same parameters ($\mu\ and\ \sigma$) [3]. The above implies that point and interval estimates for $\mu$ will be biased when the standard method of estimation is used [4]. Barendregt attempts to correct for this bias in the point estimate by using the value of the expectation of the crude odds ratio when it is assumed that the crude odds ratio follows as lognormal distribution.

Under the assumption $\widehat{OR}_{crude}$ follows a lognormal distribution with parameters ($\mu$ and $\sigma$), the formula for the expectation of a lognormally distributed random variable can be utilized:

$$x \sim LN(\mu, \sigma) \Rightarrow E(x) = exp\left(\mu + \frac{\sigma^2}{2}\right)$$

Thus, we may obtain a better estimate for $\mu$ by:

$$\hat{\mu}^* = ln(\widehat{OR}_{crude}) - \frac{\hat{\sigma}^2}{2}$$

The estimate of the parameter $\sigma$ may be obtained by the Delta method:

$$\hat{\sigma} = \sqrt{\frac{1}{a} + \frac{1}{b} + \frac{1}{c} + \frac{1}{d}}$$

(Where a, b, c and d are the cell values from the 2x2 table).

If we define $\widehat{OR}^*$ by $exp((ln(\widehat{OR}_{crude}) - \frac{\hat{\sigma}^2}{2})$, then as a means to find the sampling distribution for $\widehat{OR}^*$ consider the following:

Let: $\widehat{OR}_{crude} \sim LN(\mu, \sigma)$. This implies that $ln(\widehat{OR}_{crude}) \sim N(\mu, \sigma)$. Next, define:

$$ln(\widehat{OR}^*) = ln(\widehat{OR}_{crude}) - \frac{\hat{\sigma}^2}{2} \qquad (1)$$

Assuming that $\hat{\sigma}$ is an unbiased estimate of $\sigma$, (1) implies that:

$$ln(\widehat{OR}^*) \sim N(\mu^*, \sigma) \qquad (2)$$

Where: 
$$\mu^* = \mu - \frac{\sigma^2}{2}$$

Equation (2) follows from the fact that for a fixed finite sample size, samples of independently and identically normally distributed random variables will yield sampling distributions of the sample means that are stochastically independent from from the sample variances.

(2) implies that:

$$\widehat{OR}^* \sim LN(\mu^*, \sigma) \qquad (3)$$

It should be noted that the above method of derivation for the point estimate of $OR_{true}$, $\widehat{OR}^*$, yields a lognormal sampling distribution for this statistic, namely $\widehat{OR}^* \sim LN(\mu^*, \sigma)$, for which the parameter $\sigma$ has the same value as the corresponding parameter from the $N(\mu, \sigma)$ distribution, which is the assumed sampling distribution of $ln(\widehat{OR}_{crude})$. This is in contrast with Barendregt, who recommends recalculating the estimate of $\sigma$ based on the formula for the variance of a lognormally distributed random variable with the parameter $\mu = \mu^*$. He then recalculates the estimate of this parameter based on the new estimate of $\sigma = \hat{\sigma}^*$ to obtain final estimates $\mu = \hat{\mu}^{**}$ and $\sigma = \hat{\sigma}^*$ [4]. From equations (1), (2) and (3) and from the fact that for a random variable distributed $N(\mu, \sigma)$ the standard estimates of μ and σ are stochastically independent, the above recalculations appear to be unnecessary.

The methods used to obtain interval estimates based on these two point estimates of $OR_{true}$ are as follows:

The standard method utilizes the formula:

$$CI_\alpha = exp\left(ln(\widehat{OR}_{crude}) \pm z_{1-\alpha/2} \cdot \hat{\sigma}_{OR_{crude}}\right)$$

This confidence interval is intended to have coverage probability equal to $1 - \alpha$. In other words:

$$P(OR_{true} \in \{CI_\alpha\}) = 1 - \alpha$$

Another confidence interval may be obtained from (1) by:

$$CI_\alpha^* = exp\left(\left(ln(\widehat{OR}_{crude}) - \frac{\hat{\sigma}^2}{2}\right) \pm z_{1-\alpha/2} \cdot \hat{\sigma}_{OR_{crude}}\right)$$

The intended coverage probability for this CI is given by:

$$P(OR_{true} \in \{CI_\alpha^*\}) = 1 - \alpha$$

The mathematical justification for $CI_\alpha^*$ comes from the equations (1), (2) and (3) and the Percentile Lemma [5]. The percentile lemma implies that a confidence interval which is obtained using the parametric bootstrap method is mathematically equivalent to $CI_\alpha^*$. This method entails generating #PBS bootstrap estimates of $OR^*$ as realizations of a lognormal stochastic process with parameters $\mu^*$ and $\sigma$. The point estimate for $OR_{true}$, given by, $\widehat{OR}_{Boot}^*$ is equal to the median of the parametric bootstrap sample. The lower and upper bounds of the parametric bootstrap $\alpha$ level confidence interval are obtained as the $\alpha/2$ and $1-(\alpha/2)$ quantiles of the bootstrap sample.

Here, a comparison of the validity of these three point estimates (standard, parametric bootstrap and calculated percentile) and their associated interval estimates are obtained using Monte Carlo simulations. Additionally, the point estimate and associated confidence interval recommended by Barendregt will be included for comparison. The statistical conclusion validity of the four associated interval estimates will be assessed from the simulated empirical coverage probability [6] and confidence interval lower and upper bounds. The proportion of times that either the lower bound of the interval is greater than $OR_{true}$, or the upper bound the interval is less than $OR_{true}$, yields one minus the simulated empirical coverage probability of the CI. The proportion of times that either the lower bound of the interval is greater than 1 or the upper bound the interval is less than 1 when $OR_{true}$ is not equal to 1 out of the total number of MC simulations, will be the simulated empirical power of the two tailed hypothesis tests to reject the null hypothesis that $OR_{true}$ is equal to 1 (ie. no treatment effect), when in fact $OR_{true} \neq 1$ [6]. The theoretical power of this hypothesis test will be calculated utilizing the standard formula [7]:

$$Power = \Phi\left(|log(OR_{true})|\left[\frac{1}{n_{E=1}P(D=1|E=1)(1-P(D=1|E=1))} + \frac{1}{n_{E=0}P(D=1|E=0)(1-P(D=1|E=0))}\right]^{-1/2} - z_{\alpha/2}\right)$$

This theoretical power will be compared to the empirical power obtained from the MC simulations.

**Monte Carlo Simulation Methods:** Simulations were conducted using R [8]. The stochastic process that was utilized to generate #MC samples of $OR_{crude}$ was based on a prospective study design. The exposure status and disease outcome status were generated for a total of n=200 subjects for each simulation. Two sets of simulations were conducted, one for $OR_{true} = .279$ (exposure "protective") and one for $OR_{true} = 2.365$ (exposure "harmful"). For both examples, the exposure status E=0 (unexposed) or E=1 (exposed) were randomly assigned by a Bernoulli process with the probability of exposure set to $P(E) = .5$. Next, the probability of disease, conditional on exposure status, yielded the event probabilities for two more Bernoulli random variables whose outcomes resulted in the subject's classification in one of the cells of the two by two table. For the first example, P(D=1|E=1) = .075 and P(D=1|E=0) = .225. This yielded value for $OR_{true}$ of .279. For the second example, P(D=1|E=1) = .2667 and P(D=1|E=0) = .1333. This yielded a value for $OR_{true} = 2.365$. This process was repeated n times resulting in the complete cross-classified exposure by disease status for n subjects with cell probabilities given by:

(A) $P(D = 0, E = 0) = P(D = 0|E = 0) \cdot P(E = 0)$

(B) $P(D = 1, E = 0) = (1 - P(D = 0|E = 0)) \cdot P(E = 0)$

(C) $P(D = 1, E = 1) = P(D = 1|E = 1) \cdot P(E = 1)$

(D) $P(D = 0, E = 1) = (1 - P(D = 1|E = 1)) \cdot P(E = 1)$

For a single sample of size n, this yielded values for each of the cells of the 2 by 2 table:

|     | D=0 | D=1 |           |
| --- | --- | --- | --------- |
| E=0 | a   | b   |           |
| E=1 | c   | d   |           |
|     |     |     | n=a+b+c+d |

Where:

$E(a) = n \cdot P(D = 0, E = 0)$

$E(b) = n \cdot P(D = 1, E = 0)$

$E(c) = n \cdot P(D = 0, E = 1)$

$E(d) = n \cdot P(D = 1, E = 1)$

In order to prevent the occurrence of cells with a value of 0, .5 was added to each of the cells: a, b, c and d before the calculation of the OR estimates. For the Monte Carlo simulations, a total of #MC = 200,000 crude tables were generated. For each simulated sample, the crude odds ratio was calculated as $a \cdot d / b \cdot c$ (the standard estimate) along with the three other estimates of $OR_{true}$. Utilizing the simulated sampling distributions (#MC = 200,000), comparisons were made between the four methods of point and interval estimation of $OR_{true}$ : (I) the standard method, (II) the parametric bootstrap with a percentile CI (#PBS = 1000), (III) direct calculation of the percentile CI and (IV) the estimates proposed by Barendregt (Value in Health).

**Monte Carlo Simulation Results:**

The simulation results are presented in Table 1 and Table 2. The leftmost column (column 1) gives the point and interval estimation method used, along with the value of the point estimate of $OR_{true}$. Column 2 gives one minus the coverage probability of each 95% CI as well as the mean lower and upper bounds and the CI widths. Columns 3 and 4 give the probability that the upper bound of the CI is less than $OR_{true}$ and the probability that the lower bound of the CI is greater than $OR_{true}$ respectively (these two numbers add up to one minus the coverage probability). The rightmost column (column 5) gives the simulated empirical power of the interval based hypothesis test to reject the null hypothesis that $OR_{true} = 1$ based on the rejection criterion defined by the 95% CI not including 1.

In both Table 1 (exposure protective) and Table 2 (exposure harmful) the mean point estimates of $OR_{true}$ for the parametric bootstrap and calculated percentile methods are almost identical to $OR_{true}$. These two estimation methods also exhibit the highest level of statistical conclusion validity for their respective confidence intervals as indicated by one minus the coverage probability being close to .05. Overall, as demonstrated by the MC simulations, these two methods exhibit the least biased point estimates, the best coverage probability (as defined by it's closeness to 1-.05) and the narrowest confidence intervals of the four estimation approaches. In Table 1 (exposure protective), the two percentile based methods yielded simulation derived empirical power that was higher than the theoretical power to reject the null hypothesis that $OR_{true} = 1$ but lower than the theoretical power to reject this null hypothesis in Table 2 (exposure harmful). The simulated empirical power to reject the null hypotheses is closest to the theoretical value for the standard method of interval estimation for both cases (exposure protective or exposure harmful). The method recommended by Barendregt gives the most biased point estimates and poorest CI coverage probability in both tables. This method yielded power for the interval based hypothesis test that were midway between the standard method and the two percentile based methods for Table 1 but lower than the other three estimation methods for Table 2.

**Discussion:**

By utilizing Monte Carlo simulations, it is possible to evaluate the properties of estimators from data that are generated based on a specific stochastic process. This process can be simulated based on the statistical model from which point and interval estimates as well as hypothesis tests are formulated. Therefore, classical statistical inference, which includes both estimation and hypothesis testing, now has a means of checking if these inferential procedures are in fact valid. This type of validity has been referred to as statistical conclusion validity.

In the case of the odds ratio, there currently exist numerous alternatives for both interval estimation and hypothesis testing. This report focused on the standard method, (which is still widely used in both epidemiology and clinical trials) and three alternatives. The simple chi-square test for independence of E and D does not yield an estimate of the magnitude of the "effect" of exposure on outcome. On the other hand, the confidence interval for the odds ratio (and other relative effect measures) give the investigator a sense of how "protective" or "harmful" an exposure is. The confidence interval around the point estimate also yields a hypothesis test at the alpha level of the null hypothesis that the odds ratio equals 1 (ie, the exposure has no effect of outcome).

In this report, four different methods of point and interval estimation were examined via Monte Carlo simulations. The coverage probability was given the highest level of value as far as a measure of the statistical conclusion validity of each of the confidence interval estimation methods. By this standard, the two percentile-based methods (parametric bootstrap and calculated percentile method) yielded the highest level of validity. These intervals were also the narrowest, thus indicating a smaller amount of variability around the point estimate.

**Conclusions:** The concordance of the theoretical power of the confidence interval based hypothesis test of the null hypothesis to reject the null hypothesis that the true OR was equal to one, with the simulation derived power, was highest for the standard CI. However, for the example given of a true odds ratio that was less than one, the two percentile methods yielded the highest simulated empirical power. This result should be examined more closely, possibly with additional simulations, since if valid the higher statistical power for a protective exposure with a relatively small sample size might have particular value for researchers involved in clinical trials.

**Abbreviatons:**

OR            odds ratio

ORcrude     crude odds ratio

ORtrue       true odds ratio

ln             natural logarithm

LN           log normal probability distribution

N             normal probability distribution

CI             confidence interval

P             probability

exp          exponential

E             expectation

## Table 1 : $OR_{true} = .279$ (*Exposure Protective*)

**Intended 95 % Confidence Intervals**

Theoretical Power to reject the Null Hypothesis: $OR_{true} = 1$ is .811

**n=200** $P(E) = .5$    $P(D|E) = .075$    $P(D|\bar{E}) = .225$ **#MC = 200,000 #PBS = 1000**

| Method/ Mean point estimate | 1-P($OR_{true} \in \{CI\}$) CI: (Mean LB, Mean UB) [width] | P(miss: $OR_{true}$ too high) | P(miss: $OR_{true}$ too low) | MC Empirical Power |
|---|---|---|---|---|
| **Standard (I)**/ .309 | .040 (.131, .734) [.603] | .0091 | .0313 | .8453 |
| **Pctl Boot. (II)**/ .280 | .047 (.120, .664) [.544] | .0275 | .0199 | .8947 |
| **Pctl Calc. (III)**/ .279 | .046 (.119, .665) [.546] | .0266 | .0195 | .8945 |
| **Value in Health (IV)**/ .274 | .026 (.108, .708) [.600] | .0138 | .0124 | .8660 |

## Table 2 : $OR_{true} = 2.365$ ($Exposure\ Harmful$)

**Intended 95 % Confidence Intervals**

Theoretical Power to reject the Null Hypothesis: $OR_{true} = 1$ is .644

**n=200** $P(E) = .5$   $P(D|E) = .2667$   $P(D|\bar{E}) = .1333$ **#MC = 200,000 #PBS = 1000**

| Method/ Mean point estimate | 1-P($OR_{true} \in \{CI\}$) CI: (Mean LB, Mean UB) [Width] | P(miss: $OR_{true}$ too high) | P(miss: $OR_{true}$ too low) | MC Empirical Power |
|---|---|---|---|---|
| Standard (I)/ 2.544 | .045 (1.206, 5.394) [4.188] | .0273 | .0171 | **.6458** |
| Pctl Boot. (II)/ 2.369 | .051 (1.258, 4.991) [3.733] | .0418 | .0093 | **.5730** |
| Pctl Calc. (III)/ 2.364 | .049 (1.121, 5.006) [3.885] | .0404 | .0086 | **.5685** |
| Value in Health (IV)/ 2.338 | .037 (1.053, 5.228) [4.175] | .0330 | .0036 | **.5017** |